\documentclass[10pt]{article}

\usepackage{amsmath,amssymb,epsfig,color,bbold}
\usepackage{feynmp-auto}
\usepackage[hidelinks]{hyperref}
\usepackage{psfrag}
\usepackage{graphicx,xspace}
\usepackage{bm}
\usepackage[affil-it, auth-sc]{authblk}
\usepackage{lineno}
\usepackage[sort&compress,numbers]{natbib}
\usepackage{doi}
\usepackage{eso-pic}
\usepackage[capitalise]{cleveref}
\usepackage{comment}
\usepackage{caption}
\usepackage{subcaption}
\usepackage{slashbox}
\usepackage{physics}
\newcommand{\be}{\begin{equation}}  
\newcommand{\ee}{\end{equation}} 
\def\slash#1{#1\!\!\!/\!\,\,}  
\newcommand{\nl}{\nonumber \\ }
\renewcommand{\order}{{\cal O}}
\long\def\symbolfootnote[#1]#2{\begingroup%
\def\thefootnote{\fnsymbol{footnote}}\footnote[#1]{#2}\endgroup}

\def\dd{\mathrm{d}} 
\def\e{\mathrm{e}}

\newcommand{\iu}{{\rm i}}

\allowdisplaybreaks

\begin{document}

\begin{fmffile}{fmfnotes} 
\fmfcmd{%
vardef middir(expr p,ang) = dir(angle direction length(p)/2 of p + ang) enddef;
style_def arrow_left expr p = shrink(.7); cfill(arrow p shifted(4thick*middir(p,90))); endshrink enddef;
style_def arrow_left_more expr p = shrink(.7); cfill(arrow p shifted(6thick*middir(p,90))); endshrink enddef;
style_def arrow_right expr p = shrink(.7); cfill(arrow p shifted(4thick*middir(p,-90))); endshrink enddef;}

\fmfset{arrow_len}{2.5mm}
\fmfset{arrow_ang}{12}
\fmfset{wiggly_len}{3mm}
\fmfset{wiggly_slope}{75}
\fmfset{curly_len}{2mm}
\fmfset{zigzag_len}{2mm}

\AddToShipoutPictureFG*{\AtPageUpperLeft{\put(-60,-75){\makebox[\paperwidth][r]{FERMILAB-PUB-25-0375-T}}}}
\AddToShipoutPictureFG*{\AtPageUpperLeft{\put(-60,-60){\makebox[\paperwidth][r]{CALT-TH-2025-017}}}}

\title{\Large\bf Factorization and resummation of \\
QED radiative corrections for neutron beta decay}

\author[1]{Zehua Cao}
\author[1,2]{Richard~J.~Hill}
\author[3]{Ryan~Plestid}
\author[1,2]{Peter Vander Griend}
\affil[1]{University of Kentucky, Department of Physics and Astronomy, Lexington, KY 40506 USA \vspace{1.2mm}}
\affil[2]{Fermilab, Theoretical Physics Department, Batavia, IL 60510 USA
\vspace{1.2mm}}
\affil[3]{Walter Burke Institute for Theoretical Physics, 
California Institute of Technology, Pasadena, CA 91125 USA\vspace{1.2mm}}

\date{\today}

\maketitle

\begin{abstract}
  \vspace{0.2cm}
  \noindent
  Details of the two-loop analysis of long-distance QED radiative corrections to neutron beta decay are presented.  
  Explicit expressions are given for hard, jet, and soft functions appearing in the factorization formula that describes the small mass/large energy limit.  
  Power corrections, cancellation of singularities in the small mass expansion, renormalization scheme dependence, and bound state effects are discussed.  
  The results impact the determination of $|V_{ud}|$ from the measured neutron lifetime.  
\end{abstract}
\vfill

\newpage
\tableofcontents

\vfill
\pagebreak

\begin{figure}[htb]
\centering
\includegraphics[width=0.6\textwidth]{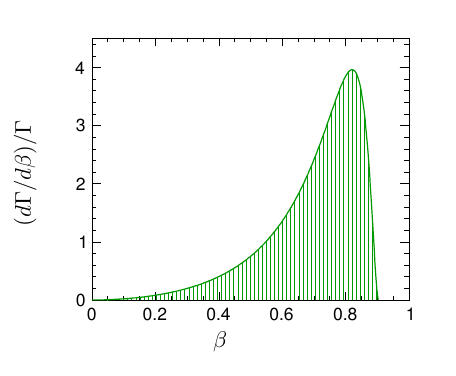}
\caption{Electron velocity spectrum for neutron beta decay at tree level. The spectrum is dominated by $\beta\gtrsim 0.5$, which causes non-relativistic approximations (such as the Fermi function) to converge slowly. \label{fig:betaspec}}
\end{figure}

\bigskip

\section{Introduction}\label{sec:intro}

Low-energy processes, such as neutron and nuclear beta decay, are critical for determining the fundamental constants of Nature~\cite{Bopp:1986rt,Ando:2004rk,Darius:2017arh,Seng:2018yzq,Seng:2018qru,Fry:2018kvq,Czarnecki:2019mwq,Hayen:2020cxh,Seng:2020wjq,Gorchtein:2021fce,UCNt:2021pcg,Shiells:2020fqp}. These processes currently dominate the determination of the quark mixing parameter $|V_{ud}|$ and serve as fruitful testing grounds for physics beyond the Standard Model~\cite{Cirigliano:2013xha,Glick-Magid:2016rsv,Gonzalez-Alonso:2018omy,Glick-Magid:2021uwb,Falkowski:2021vdg,Brodeur:2023eul,Crivellin:2020ebi,Coutinho:2019aiy,Crivellin:2021njn,Crivellin:2020lzu,Cirigliano:2022yyo}. At current levels of experimental precision, control over theoretical predictions is required at the 100 ppm (i.e., $\sim 10^{-4}$) level. 

\bigskip

Quantum electrodynamic (QED) radiative corrections are required in order to interpret experimental results as measurements of fundamental constants. At the current $\sim 10^{-4}$ experimental precision \cite{Hardy:2020qwl}, two-loop
(i.e., $O(\alpha^2)$ in the fine structure constant) 
corrections to neutron beta decay are relevant. It is well known that at one-loop order the neutron beta decay rate receives an ``unnaturally large'' radiative correction scaling as $\pi\alpha/\beta$ with $\beta$ the electron velocity~\cite{Wilkinson:1982hu}. 
Such a large first-order correction 
calls into question the rate at which QED radiative corrections converge, and the possible relevance of yet higher-order (e.g., three-loop) corrections. 
With this motivation, we examine the origin and description of enhanced contributions to the neutron decay 
rate to all orders, using renormalization group methods.

\bigskip 

The large first order QED radiative correction to {\it neutron} beta decay has historically been associated with the Fermi function for {\it nuclear} beta decay~\cite{Fermi:1934hr,Wilkinson:1982hu,Czarnecki:2004cw}. The Fermi function, however, is justified only for large nuclear charge $Z\gg 1$ or small electron velocity $\beta\ll 1$.  Clearly the former condition is not satisfied in neutron decay ($Z=0$ or $Z=1$ for the neutron or proton).  As \cref{fig:betaspec} illustrates, the electron spectrum is also strongly suppressed at small velocity.  In Ref.~\cite{VanderGriend:2025mdc}, we demonstrated how the opposite, $m\to 0$, limit provides a controlled expansion.  
Here we discuss the factorization formalism underlying this analysis and provide explicit two-loop results for the components of the factorization formula.

\vfill
\pagebreak

\section{Soft-hard factorization}

Neutron beta decay is described by a low-energy effective Lagrangian involving spin-1/2 heavy particle fields:~\cite{Hill:2023acw}
\begin{align}\label{eq:neutronL}
  {\cal L}_{\rm eff} &=  -\bar{h}_v^{(p)}\left( {\cal C}_V \gamma^\mu +{\cal C}_A \gamma^\mu \gamma_5\right) h_v^{(n)}
  \bar{e} \gamma_\mu (1-\gamma_5) {\nu}_e + {\rm H.c.} \,,
\end{align}
where $h^{(n)}_{v}$ and $h^{(p)}_{v}$ are heavy-particle spinors that annihilate neutrons and protons of velocity $v$, with $v^\mu=(1,0,0,0)$ in the laboratory frame. 
This Lagrangian corresponds to the leading order in an expansion around the static limit in which the nucleon masses are taken to infinity while their mass difference, $\Delta = m_{n} - m_{p}$, is fixed.
Higher-order terms in the expansion encode recoil corrections and are suppressed by powers of the inverse nucleon mass $M = (m_{n}+m_{p})/2$; the effective theory Lagrangian can be easily extended to include such corrections~\cite{Ando:2004rk}.
The scheme dependent Wilson coefficients 
$\mathcal{C}_{V,A} \approx(G_F/\sqrt{2}) \, V_{ud} \, g_{V,A}$ encode hadronic structure
and are determined by experimental measurement and/or nonperturbative 
QCD matching calculations (e.g. chiral perturbation theory or lattice QCD). 

The QED corrections discussed in the Introduction originate from long-distance scales and can be computed in the effective theory described by Eq.~\eqref{eq:neutronL}. 
The separation of the hadronic scale ($\Lambda_{\rm UV} \sim m_\pi \sim 100\,{\rm MeV}$) from kinematic scales of neutron beta decay 
($p \sim \Delta \sim m \sim {\rm MeV}$), and 
the soft scale ($\lambda \sim \varepsilon_\gamma \ll p$, where $\lambda$ and $\varepsilon_\gamma$ denote photon mass regulator and photon energy cutoff parameter, respectively) 
imply the factorization of the decay amplitude schematically as 
\begin{align}\label{eq:full_factorization}
    {\cal M} \sim {\cal M}_{\rm UV}( \Lambda_{\rm UV} / \mu_{\rm UV}) {\cal M}_H(p/ \mu_{\rm UV}, p/\mu) {\cal M}_S(\lambda/\mu) \,,
\end{align}
where $\mu_{\rm UV}$ and $\mu$ are factorization scales separating the UV, hard, and soft regions.  
For an electron with momentum $p$ and a neutrino with momentum $k$, we write the 
complete amplitude as 
\begin{align}
 {\cal M} &= - \bar{u}_v \left( {\cal C}_V \gamma^\mu +{\cal C}_A \gamma^\mu \gamma_5\right)u_v  \, \bar{u}(p) \mathcal{M}_H \mathcal{M}_S \gamma_\mu (1-\gamma_5) v(k) \,,
\end{align}
where $u_v$ is the heavy particle spinor wavefunction. 
The UV amplitude ${\cal M}_{\rm UV}$ is thus given simply in terms of ${\cal C}_V$ and ${\cal C}_A$, and the soft amplitude ${\cal M}_S$ exponentiates~\cite{Hill:2023acw,VanderGriend:2025mdc}.  In the following subsections, we discuss the hard amplitude ${\cal M}_{H}$.

\subsection{One loop computation}\label{subsec:one_loop}
Let us study the one-loop corrections to neutron beta decay in order to better understand the origin of the 
term proportional to $\pi \alpha/\beta$.
In what follows, we denote the amplitude with a proton in the final state by ${\cal M}_H(w, \mu^{2})$, whereas the crossed amplitude is denoted by $[{\cal M}_H]_{v\rightarrow -v}$.  
As we will see in what follows, the large ``$\pi$-enhanced" radiative correction is present in ${\cal M}_H(w, \mu^{2})$ but absent in  $[{\cal M}_H]_{v\rightarrow -v}$.  
We trace the origin of this term to an infrared divergence in the hard amplitude, which is in turn related to an ultraviolet divergence in the soft effective theory, where the electron is also treated as a heavy particle.  Such terms can therefore be handled with renormalization group techniques. 

The Lagrangian (\ref{eq:neutronL}) implies that, to all orders in perturbation theory, the hard function ${\cal M}_H$ takes the form (suppressing $\mu_{\rm UV}$ dependence)
\begin{align}
{\cal M}_H(w, \mu^{2}) = {\cal A}_H(w, \mu^{2}) + \frac{1}{w} \slash{v} {\cal B}_H(w, \mu^{2}) \,,
\end{align}
for scalar functions ${\cal A}_H$ and ${\cal B}_H$,
where 
$w = v\cdot p / m$, and $p^{\mu}$ and $m$ are the electron four-momentum and mass.
A straightforward one loop computation, including wavefunction renormalization factors, yields
\begin{align}\label{eq:ah_bh}
        {\cal A}_H(w) &= 1 + {\alpha \over 2\pi}\bigg[ \frac34 \log{\mu_{\rm UV}^2\over m^2} 
    + \log{\mu^2\over m^2} ( \tilde{w} j(\tilde{w}) - 1) 
    + \tilde{w}j(\tilde{w}) 
    - \tilde{w}J(\tilde{w}) 
    \bigg] \,,
    \\
    {\cal B}_H(w) &= {\alpha \over 2\pi}\left[ - \tilde{w} j(\tilde{w}) \right] \,,
\end{align}
where $\tilde{w} = -w - \iu0$ and the functions $j$ and $J$ are defined by
\begin{align}
    xj(x) &= \int_0^\infty dy\, \frac{x}{1 + y^2 + 2yx}
     \,,
   \nl
    xJ(x) &=  \int_0^\infty dy\, \frac{x}{1 + y^2 + 2yx} \log(1 + y^2 + 2yx) \,.
\end{align}
The term $\alpha/(2\pi) \times \frac34 \log(\mu^2_{\rm UV}/m^2)$ is related to the renormalization group flow of the Wilson coefficients $\mathcal{C}_{A,V}$, or equivalently $\mathcal{M}_{\rm UV}$, as denoted explicitly in \cref{eq:full_factorization}.

Let us consider the related quantity $[{\cal M}_{H}]_{v\to -v}$, 
which describes the spacelike decay of a heavy negatively charged particle decaying into an electron.
Using, for $w>1$,
\begin{align}
    wj(w) 
    &= \frac{w}{\sqrt{w^{2}-1}} \log{(w+\sqrt{w^{2}-1})}
     \,, \nl
    wJ(w) 
    &= \frac{w}{\sqrt{w^{2}-1}} \left( {\rm Li}_{2}(1 - (w - \sqrt{w^{2}-1})^{2}) + \log^{2}{(w+\sqrt{w^{2}-1})} \right) \,,
\end{align}
we find 
\begin{align}
        {\cal A}_H(-w) &= 1 + {\alpha \over 2\pi}\bigg[ \frac34 \log{\mu_{\rm UV}^2\over m^2} 
    + \log{\mu^2\over m^2} ( w j(w) - 1) 
    + wj(w) 
    - wJ(w) 
    \bigg] \,,
    \nl
    {\cal B}_H(-w) &= {\alpha \over 2\pi}\left[ - w j(w) \right] \,.
\end{align}
To recover the timelike process corresponding to neutron beta decay, 
we use that 
\begin{equation}
\begin{split}
    \tilde{w}j(\tilde{w}) =& wj(w) - \iu\pi \frac{w}{\sqrt{w^{2}-1}},\\
    \tilde{w}J(\tilde{w}) =& w J(w) - \iu\pi\frac{w}{\sqrt{w^{2}-1}} \log(-4(w^{2}-1)-\iu0).
\end{split}
\end{equation}
We note that the spacelike amplitude is finite at $w\to 1$
\begin{align}
     \lim_{w\to1} {\cal A}_H(-w) &= 1 + {\alpha \over 2\pi}\bigg[ \frac{3}{4} \log{\mu_{\rm UV}^2\over m^2} -1
    \bigg] \,,
    \nl
    \lim_{w\to1} {\cal B}_H(-w) &= {\alpha \over 2\pi} ( - 1 )\,.
\end{align}
In contrast, the timelike amplitude contains an enhancement 
$\log(-1-\iu0) \log[(-4\bm{p}^{2}-\iu0) / \mu^{2}] \sim -\pi^{2}$
(at $\mu^2 = 4\bm{p}^2$), and diverges at $w\to 1$: 
\begin{align}\label{eq:1loopMH}
    {\cal A}_H(w)-{\cal A}_H(-w) &= {\alpha\over 2\pi} \bigg[ 
    {\iu\pi w\over\sqrt{w^2-1}}\left(  \log\left( -4\bm{p}^2-\iu0 \over \mu^2  \right) -1  \right) 
    \bigg] \,,
    \nl
    {\cal B}_H(w) - {\cal B}_H(-w) &= {\alpha\over 2\pi} \bigg[  {\iu\pi w\over\sqrt{w^2-1}} \bigg] \,.
\end{align}
Evaluating at $\mu^{2} = \hat{\mu}^2 = -4\bm{p}^{2}-\iu 0$
minimizes such large logarithms; equivalently, 
enhancements present in the timelike amplitude can be obtained 
with a renormalization group resummation
\begin{equation}
    {\cal M}_{H}(w, \mu^{2}) = {\rm exp}\left[ \int_{\hat{\mu}}^{\mu} {d\mu^{\prime } \over \mu^{\prime }} \,\, \gamma_{H}\right] {\cal M}_{H}(w, \hat{\mu}^{2})\,,
\end{equation}
where $\gamma_{H}$ is the anomalous dimension of the hard coefficient.

\vfill
\pagebreak

\subsection{Anomalous dimension and resummation of large logarithms}

Since ${\cal M}_H$ represents the coefficient function of a soft operator, 
its $\mu$ dependence is given, to all orders in perturbation theory, by the cusp anomalous dimension~\cite{Korchemsky:1987wg,Korchemskaya:1992je}, 
\begin{align}
\gamma_H 
= -\frac{{\rm d}\log{\cal M}_S(\mu^{2})}{{\rm d}\log\mu} 
 = {\alpha \over \pi} \left[ -1 + {1\over 2\beta}\log{1+\beta\over 1-\beta} - {\iu\pi \over \beta} \right], 
\end{align}
with solution
\begin{align}\label{eq:MH_mu_evolution}
    {\cal M}_H(\mu^2)
    &= \exp\bigg[\frac{\alpha}{2\pi}\left( -1 + {1\over 2\beta}\log{1+\beta\over 1-\beta}    \right)\log{\mu^2 \over \hat{\mu}^2}  \bigg]
    \exp\bigg[{-\iu\alpha \over 2\beta }\log{\mu^2\over \hat{\mu}^2}   \bigg]
    {{\cal M}_H(\hat{\mu}^2)}     .
\end{align}
For $\mu^2= 4\bm{p}^2$ and $\hat{\mu}^2=-4\bm{p}^2-\iu0$, we have 
$\log{\mu^2 \over \hat{\mu}^2} = +\iu\pi$. 
The first exponential in \cref{eq:MH_mu_evolution} is then a phase that does not 
impact rate observables and the second exponential
leads to a numerical enhancement for the associated hard function,
\footnote{
The hard function is given by the appropriate average of the squared magnitude of ${\cal M}_H$.  For example, in the electron energy spectrum for neutron decay  (cf. Eq.~(14) or Ref.~\cite{VanderGriend:2025mdc}), 
$H(\mu^{2}) = \langle \, |{\cal M}_{H}(\mu^{2})|^{2} \, \rangle$,  
where the angle brackets denote contracting with lepton and nucleon spinors, 
$\mathcal{M}_H \to 
\bar{u}(p) \mathcal{M}_H \gamma_\mu (1-\gamma_5) v(k) 
\bar{u}_v \left( {\cal C}_V \gamma^\mu +{\cal C}_A \gamma^\mu \gamma_5\right) u_v$, 
squaring, summing over final state spins and integrating over final state phase space (apart from the electron energy), 
and dividing by the same expression at tree level. 
}
\begin{align}\label{eq:Hratio}
    H(\mu^2)
    =  \exp\bigg[{ \pi \alpha \over \beta }   \bigg] 
   H(\hat{\mu}^2)
    \,.
\end{align}
We emphasize that this result is true to all orders and makes no reference to non-relativistic kinematics. 
Compared to the expression involving the naive application of the traditional Fermi function for nuclear beta decay, the expression (\ref{eq:Hratio}) differs starting at two-loop order (cf.  Sec.~6 of Ref.~\cite{VanderGriend:2025mdc}).

\section{Soft-collinear-hard factorization}

Here we analyze the large-energy/small-mass limit characterized by $E\gg m$. 
This is sufficient to compute the two-loop corrections for neutron beta decay to the requisite $10^{-4}$ precision, and provides an all-orders derivation of the result (\ref{eq:Hratio}) in this limit.

\subsection{Factorization formula}

Let us decompose the hard function in \cref{eq:full_factorization}, neglecting contributions suppressed by powers of $m/E$, 
\begin{align}
{\cal M}_H(w) \approx {\cal A}_H(w) \approx 
F_R(w,m) F_J(m) F_H(E) \,. 
\end{align}
Power corrections are discussed in Appendix~\ref{app:power}. 
We will analyze the remainder $F_R$, jet $F_J$ and hard $F_H$ amplitudes, using dimensional regularization and the $\overline{\rm MS}$ scheme.   
Following the discussion in the previous section, we consider first the quantity ${\cal M}_H | _{v\to -v} 
\approx {\cal A}_H(-w) = F_R(-w,m) F_J(m) F_H(-E)$, and then analyze the ratio ${\cal A}_H(w)/{\cal A}_H(-w)$ using renormalization arguments to resum enhanced contributions. 
Note that the product $F_S F_R F_J F_H$ is equal to the
heavy-light Dirac form factor with heavy-particle initial state, at leading power.
The renormalized soft amplitude is given, to all orders in the coupling, as 
\begin{align}
    F_S &= \exp\left[ {\bar{\alpha}_0 \over 4\pi} \left(
    2(L-1) \log{\lambda^2\over \mu^2} 
    \right)\right] \,,
\end{align}
where $L\equiv \log(2E/m)=\log(2w)$ and $\bar{\alpha}_0 \to \alpha$ is the $\overline{\rm MS}$ coupling for $n_{e}=0$, which reduces to the on-shell coupling in $d=4$. 

\subsection{Remainder function}

The quantity $F_R$ represents the matching coefficient between the soft operator 
(Wilson line operator for charge $-1$ particles 
involving initial state $v$ and final state $v^\prime$)
in $n_{e}=1$ and $n_{e}=0$.   This correction begins at two loop order, necessarily involving a loop of the massive lepton. 
From \cref{app:remainder}, the bare result is 
\begin{align}
F_R^{\rm bare}(-w,m)
    &=
Z_{h, n_e=1}\Bigg[\quad
   \parbox{30mm}{
\begin{fmfgraph*}(75,85)
  \fmfleftn{l}{3}
  \fmfrightn{r}{3}
  \fmftopn{t}{3}
  \fmf{phantom}{l2,v,r2}
  \fmffreeze
  \fmf{double}{r2,x,v}
  \fmf{double}{l2,y,v}
\fmfv{decor.shape=circle,decor.filled=full,decor.size=2mm}{v}
\end{fmfgraph*} 
}
+ \quad
    \parbox{35mm}{
\begin{fmfgraph*}(100,85)
  \fmfleftn{l}{3}
  \fmfrightn{r}{3}
  \fmftopn{t}{3}
  \fmf{phantom}{l2,v,r2}
  \fmffreeze
  \fmf{double}{r2,x,v}
  \fmf{double}{l2,y,v}
  \fmffreeze
  \fmf{fermion,left,tension=0.25}{a1,a2,a1}
  \fmf{phantom}{t2,a1}
  \fmf{phantom}{t2,a2}
  \fmf{phantom}{x,a1}
  \fmf{phantom}{a2,y}
  \fmffreeze
\fmfv{decor.shape=circle,decor.filled=full,decor.size=2mm}{v}
\fmf{photon,right=0.5}{x,a1}
\fmf{photon,right=0.5}{a2,y}
\end{fmfgraph*} 
}
+ \dots \quad \Bigg]
\nonumber  
    \\[-10mm] 
    &=1+
    \left( {\bar{\alpha}_{1}\over 4\pi} \right)^2 \left({m^2\over \mu^2}\right)^{-2\epsilon} 
 n_{e}\left(
 -{4\over 3\epsilon^2} + {20\over 9\epsilon} - {112\over 27} - {2\pi^2 \over 9}
  + \order(\epsilon)
 \right)
 (L - 1)
 \,,
\end{align}
where $Z_{h,n_e=1}$ is the on-shell wavefunction renormalization factor for $n_e=1$, and we have used 
$wj(w) \to L$ in the large energy limit (with $L=\log(2E/m)$ as defined above). 
The renormalized remainder function is then given by 
\begin{align}
    F_R(-w,m,\mu) &= Z_S Z_{S, n_{e}=1}^{-1} F_R^{\rm bare}(-w,m) \,,
\end{align}
where the soft ($n_{e}=0$) operator renormalization is 
\begin{align}
    Z_S &= \exp\left[ {\bar{\alpha}_0\over 4\pi}{2\over \epsilon}(-L + 1) \right]~.
\end{align}
The soft operator renormalization for $n_{e}=1$ is related to the cusp anomalous dimension, and through two loops is~\cite{Korchemsky:1987wg,Kilian:1992tj}
\begin{align}
    Z_{S, n_{e}=1} &= 1 + {\bar{\alpha}_1 \over 4\pi} {2\over \epsilon}(-L+1) 
    + \left({\bar{\alpha}_1 \over 4\pi}\right)^2
    \bigg\{ 
    {2\over \epsilon^2}(-L+1)^2 + n_{e} 
    \left(-{4\over 3\epsilon^2} + {20\over 9\epsilon} \right)(L-1) 
    \bigg\} \,.
\end{align}
The quantities $\bar{\alpha}_0$ and $\bar{\alpha}_1$
denote the $\overline{\rm MS}$ couplings with 
$n_{e}=0$ and $n_{e}=1$ dynamical electrons, respectively; they
are related by 
\begin{align}
    \bar{\alpha}_1 &= \bar{\alpha}_0 \left[ 
    1 + n_{e} {\bar{\alpha}_0 \over 4\pi}\frac43 \left(- {1\over \epsilon}~+ \e^{\gamma_E \epsilon} \Gamma(\epsilon) \left(m^2\over \mu^2\right)^{-\epsilon}  \right)
    \right]
    \nl
    &= \bar{\alpha}_0 \left\{
    1 + n_{e} {\bar{\alpha}_0 \over 4\pi}
    \left[ - \frac43 \log{m^2\over \mu^2}
    + \epsilon\left(
    \frac23 \log^2{m^2\over \mu^2} + {\pi^2 \over 9}
    \right) + \order(\epsilon^2) \right]
    \right\} \,.
\end{align}
Substituting, we obtain
\begin{align}\label{eq:FRmu}
    F_R(-w,m,\mu) &= 1 + \left(\bar{\alpha}_1 \over 4\pi \right)^2 (L-1) n_{e} 
    \left( -\frac43 \log^2{m^2\over \mu^2} - \frac{40}{9}\log{m^2\over \mu^2} - \frac{112}{27} \right) \,.
\end{align}

\subsection{Jet function}

We have extracted the two-loop jet function using $F_R$ and results from Ref.~\cite{Hill:2016gdf} (see \cref{app:jet-function} for details), 
\begin{align}\label{eq:FJmu}
    F_J(m, \mu) &= 1 + {\bar{\alpha}_1\over 4\pi}
    \left[ \frac12 L_m^2 -\frac12 L_m + 2 + {\pi^2\over 12} \right]
\nl
&\quad +  \left({\bar{\alpha}_1\over 4\pi}\right)^2\bigg[ 
    \frac18 L_m^4 
   -\frac16 \left( - \frac43 n_{e} + \frac32 \right) L_m^3 
    -\frac14\left( \frac{52}{9} n_{e}  - \frac92 - {\pi^2\over 6} \right) L_m^2 
    \nl
    &\quad 
    -\frac12 \left( n_{e}\left( -{154\over 27} - {8\pi^2\over 9}\right)
    + \frac72 - {23\pi^2\over 12} + 24\zeta_3
    \right)L_m
    \nl
    &\quad 
+ n_{e} \left( {4435\over 324} - \frac29 \zeta_3 -{41\pi^2\over 54}\right)
   + 2\pi^2 -4\pi^2\log{2} - {331\pi^4\over 1440} - 3\zeta_3 + {209\over 16}
    \bigg] \,,
\end{align}
where $L_m \equiv \log(m^2/ \mu^2)$.

\subsection{Hard function}

As described in \cref{app:hard_function}, 
the hard function is given by $F_H=F_1/C(\mu)$, where $F_1$ 
denotes the relativistic Dirac heavy-light form factor computed with vanishing light charged fermion mass, and $C$ denotes the matching coefficient from relativistic to effective heavy particle treatment of the heavy charged fermion. 
Using the results for $F_1$ from Ref.~\cite{Beneke:2009rj} 
(see also Ref.~\cite{Bonciani:2008wf}) 
and the results for $C$ from Refs.~\cite{Broadhurst:1994se, Grozin:1998kf, Bekavac:2009zc}, 
the renormalized hard amplitude is
\begin{align}
    F_H(-E) &= 1 +  {\bar{\alpha}_1\over 4\pi}
    F_H^{(1)}(-E)
    + \left( {\bar{\alpha}_1\over 4\pi}\right)^2 F_H^{(2)}(-E) \,,
\end{align}
with 
\begin{align}
    F_H^{(1)}(-E,\mu) &= -2 L_E^2  + 2 L_E - 2 - {5\pi^2 \over 12} \,,
    \nl
    F_H^{(2)}(-E,\mu) &=
    n_{e}\bigg[ -\frac{16}{9}L_E^3 + \frac{64}{9}L_E^2 + \left(-\frac{304}{27}-\frac{16\pi^2}{9} \right)L_E
    +{656\over 81} + \frac29\zeta_3
    + {113\pi^2 \over 54}
    \bigg] \nl
    &\quad
    + 2L_E^4 - 4L_E^3 
    + \left(6 + {5\pi^2\over 6}\right)L_E^2
    + \left( 24\zeta_3- {11\pi^2\over 2}\right) L_E - 8 + {65\pi^2\over 6}
    - {167\pi^4\over 288} - 15 \zeta_3 \,,
\end{align}
where $L_E \equiv \log(2E/\mu)$.
 
Setting $\mu_{\rm UV}=\mu$ in both the Wilson coefficient and the hard function, it is readily seen that the product $C(\mu) F_H(\mu)$ has the expected scale variation for the coefficient of an effective operator with one heavy and one light energetic field~\cite{Hill:2016gdf,Becher:2003kh,Becher:2009kw,Becher:2009qa,Beneke:2009rj}, 
\begin{align}
    \label{RG-C-FH}
       {{\rm d} \log C(\mu) F_H(-E,\mu) \over {\rm d}\log\mu} &= 
       -{{\rm d} \log (F_J F_R F_S) \over {\rm d}\log\mu} 
       \nl &=
\gamma^h(\bar{\alpha}) + \gamma^\psi(\bar{\alpha}) 
    + \gamma_{\rm cusp}(\bar{\alpha}) L_{E}  
    \nl
    &=       {\bar{\alpha} \over 4\pi} \left( 4 L_E - 5 \right)
+ \left( {\bar{\alpha} \over 4\pi} \right)^2 
\left[ 
 n_{e} \left(  -{80\over 9} L_E + {250\over 27} + {2\pi^2\over 3} \right) - \frac32 + 2\pi^2 - 24\zeta_3 
\right] + \dots \,,
\end{align}
where $\gamma^{h}$ and $\gamma^{\psi}$ are the one-body
heavy-particle and light-particle contributions, 
and $\gamma_{\rm cusp}$ is the cusp anomalous dimension contribution.  

\section{Renormalization}

Enhanced contributions result from the continuation $F_H(-E) \to F_H(E)$.  To see the all-orders properties of this continuation, we recognize that  (again setting $\mu_{\rm UV}=\mu$) $F_H$
depends only on the dimensionless ratio $E/\mu$, and that
the $\mu$ dependence is determined by renormalization.  In particular, accounting for the $\mu$ dependence of $C(\mu)$, the hard amplitude obeys
\begin{align}\label{eq:twoloopanom}
       {{\rm d} \log F_H(E,\mu) \over {\rm d}\log\mu} &= 
       - \gamma_{\rm UV} + 
\gamma^h(\bar{\alpha}) + \gamma^\psi(\bar{\alpha}) 
    + \gamma_{\rm cusp}(\bar{\alpha}) \tilde{L}_{E}  
    \nl
    &=       {\bar{\alpha} \over 4\pi} \left( 4 \tilde{L}_E - 2 \right)
+ \left( {\bar{\alpha} \over 4\pi} \right)^2 
\left[ 
 n_{e} \left(  -{80\over 9} \tilde{L}_E + {160\over 27} + {2\pi^2\over 3} \right) - 4 + {14\pi^2\over 3} - 24\zeta_3 
\right] + \dots \,,
\end{align}
with $\tilde{L}_{E} = \log[(-2E-\iu0)/\mu]$.
Here we integrate this equation to obtain 
\begin{align} \label{eq:FHresum}
    {F_H(E,\mu)\over F_H(-E,\mu)} =  {F_H(E,\mu)\over F_H(E,-\mu-\iu0)}
  \,.
\end{align}

\subsection{Coupling constant renormalization}

To compute the solution to \cref{eq:FHresum}, let us consider
\begin{align}
 X \equiv 2 \log{\mu \over \hat{\mu}} = 
 2\int_{\hat{\alpha}}^\alpha {d\alpha \over \beta(\alpha)} \,,
\end{align}
for renormalization scales $\mu$ and $\hat{\mu}$,  
where
\footnote{The leading coefficients of the QED beta function are $\beta_0=-\frac43 n_e$, $\beta_1 = -4 n_e$. }
\begin{align}
    \beta = {\rm d}\alpha/{\rm d}\log\mu 
    = -2\alpha \left[ \beta_0 a + \beta_1 a^2 + \dots \right] \,,
\end{align}
and we write $a=\alpha/(4\pi)$.  
We readily obtain
\begin{align}
\hat{a}^{-1} &= a^{-1} - \beta_0 X - {\beta_1\over \beta_0} \log{\hat{a}\over a} -\left( {\beta_2\over \beta_0} - {\beta_1^2\over \beta_0^2} \right)(\hat{a} - a) + \dots \,.
\end{align}
Before expansion in $aX$, the solution expanded in $a$ is 
\begin{align}\label{eq:rexp}
    r = {a\over \hat{a}} &= (1-a\beta_0 X)+a{\beta_1\over\beta_0}\log(1-a\beta_0 X) 
    \nl & \quad 
    + a^2\bigg[
    {\beta_1^2\over\beta_0^2}(1-a\beta_0 X)^{-1}\log(1-a\beta_0 X)
    -\left(  {\beta_2\over \beta_0} - {\beta_1^2\over \beta_0^2}  \right){a\beta_0 X\over 1-a\beta_0 X}
    \bigg] + \dots \,.
\end{align}

\subsection{Coefficient renormalizaion}

In terms of $r=a/\hat{a}$, and at $\hat{\mu}=\hat{\mu}_* = -2E-\iu0$, 
\begin{align}
    \log{F_H(E,\mu)\over F_H(E,\hat{\mu}_*)}
    &= -{\gamma_0 \over 2\beta_0} \left[
    \log{r} + a\left(1-r^{-1}\right) \left({\gamma_1\over \gamma_0} - {\beta_1\over \beta_0}\right) + \dots \right]
    \nl
    &\quad 
    -{\gamma_{\rm cusp}^{(0)}\over 4\beta_0^2} \bigg\{
{1\over a} \left( 1 - r + r\log{r}\right) 
+ \left( {\gamma_{\rm cusp}^{(1)}\over\gamma_{\rm cusp}^{(0)}}-{\beta_1\over\beta_0} \right)(-\log{r}+r-1)
-{\beta_1\over 2\beta_0}\log^2{r}
\nl &\quad
+ a{1\over 2r}\bigg[
{\beta_2\over \beta_0}(1-r^2+2\log{r}) +{\beta_1^2\over\beta_0^2}(1-r)(1-r-2\log{r})
\nl
&\quad
+ {\beta_1\over \beta_0}{\gamma_{\rm cusp}^{(1)}\over\gamma_{\rm cusp}^{(0)}}(-3+4r-r^2-2r\log{r}) + {\gamma_{\rm cusp}^{(2)}\over\gamma_{\rm cusp}^{(0)}}(1-r)^2
\bigg] +
\dots \bigg\} \,,
\end{align}
with $r$ given by \cref{eq:rexp} and $X=2\log[\mu/(-2E-\iu0)] = -2 \tilde{L}_E  = 2\pi\iu -2 L_E$.\footnote{The leading anomalous dimension coefficients are 
(cf.~\cref{RG-C-FH,eq:twoloopanom} and references above)
$\gamma_0 = -2$, $\gamma_1 = \left( \frac{2\pi^2}{3} + \frac{160}{27} \right) n_e + \frac{14\pi^2}{3} - 24\zeta_3 - 4$, $\gamma_{\rm cusp}^{(0)} = 4$, $\gamma_{\rm cusp}^{(1)} = \left(-\frac{80}{9}\right) n_e$.
}

Let us focus on the large logarithm represented by $X_* = X|_{\mu=2E} = 2\pi\iu$.  Counting $|X|^4 \sim a^{-1}$, we may expand order by order including these logarithmic enhancements.
In particular, at $\mu=2E$, $X=X_*$ is pure imaginary, and 
\begin{align}
    \label{RG-enhancement}
    \left| { F_{H}(\mu_*) \over F_{H}(\hat{\mu}_{*})} \right|^2
    &= \exp\bigg[ -X_{*}^2 a + \frac{32}{9}n_{e} X_{*}^2  a^2 -\frac{8}{27} n_{e}^2 X_{*}^4 a^3
    + \dots
    \bigg] 
    \nl
    &\approx \left[ 1 + 9.9\,{\bar{\alpha}\over \pi} +
    48.7  \left(\bar{\alpha}\over \pi\right)^2
+ 160.2 \left(\bar{\alpha}\over \pi\right)^3 
+ \order(\alpha^4)
\right]
\nl
&\hspace{0.1\linewidth}\times \left[ 1 - 8.8 n_e \left({\bar{\alpha}\over \pi}\right)^2 + \order(\alpha^4) 
\right]  \times 
\left[ 1 - 7.2 n_e^2 \left({\bar{\alpha}\over \pi}\right)^3 + \order(\alpha^4) 
\right] 
    \,.
\end{align}
In the final line, we illustrate the enhanced coefficients as a series in $\bar{\alpha}/\pi$, where we have separated the series into expansions involving $X_*^2a$, $X_*^2 a^2$, and $X_*^4a^3$. 

\subsection{Hard function without large logarithms}
Choosing $\hat{\mu} = -2E-\iu0$ yields $\log[(-2E-\iu0)/\hat{\mu}] =0$, and  
the hard function is then 
\begin{align}
    F_H(\hat{\mu}) &= 1 +  {\bar{\alpha}_1(\hat{\mu})\over 4\pi}
    F_H^{(1)}(\hat{\mu})
    + \left( {\bar{\alpha}_1(\hat{\mu})\over 4\pi}\right)^2 F_H^{(2)}(\hat{\mu}) \,,
\end{align}
with 
\begin{align}\label{eq:FH}
    F_H^{(1)}(\hat{\mu}) 
    &= -2 -\frac52 \zeta_2
    \,,
    \nl
    F_H^{(2)}(\hat{\mu})
    &= n_{e}\bigg[ 
     {656\over 81} + \frac{113}{9} \zeta_2 + \frac29\zeta_3
    \bigg] 
    - 8 + 65 \zeta_2 - 15 \zeta_3 - \frac{835}{16}\zeta_4
    \,.
\end{align}
Using that 
\begin{align}
    \hat{a}_* = a - a^2 \beta_0 X_* + \order(a^3 X^2) \,,
\end{align}
with $X_*$ pure imaginary, we may ignore corrections to $\hat{a}/a$ when computing the 
modulus-squared coefficient, 
\begin{align}
    \label{order-1-numbers}
        |F_H(\hat{\mu})|^2 &= 1 - 3.1 \left({\bar{\alpha}\over \pi}\right) + (5.4+ 3.6n_e) \left(\bar{\alpha}\over \pi\right)^2 + \dots \,,
\end{align}
where $\bar{\alpha} =\bar{\alpha}_1(\mu=2E)$. 
We observe the expected, order unity, coefficients, to be compared with the enhancements that are resummed above. Had we instead chosen to use $\mu$ instead of $\hat{\mu}$ (i.e., combining \cref{RG-enhancement,order-1-numbers}) we would instead obtain, 
\begin{equation}
    |F_H(\mu)|^{2} = 1 + 6.8 \left({\bar{\alpha}\over \pi}\right) + (23.9-5.1 n_e) \left(\bar{\alpha}\over \pi\right)^2 + \dots \,,
\end{equation}
which serves to illustrate how factors of $X_*^2$ can slow the convergence of perturbation theory. The numerical values for subleading perturbative coefficients are scheme dependent. Different scheme choices, and the relations between them, are discussed in \cref{app:ren-schemes}.

\section{Higher-order corrections to the neutron lifetime}
We have carried out a calculation of long-distance radiative corrections to the neutron lifetime at $O(\alpha^2)$, identifying the well-known numerical enhancement at first order in perturbation theory as a large logarithm and performing a renormalization analysis to systematically resum such logarithms to all orders.  
In this section, we analyze $\beta\to 0$ singularities appearing at ${\cal O}(\alpha^3)$ and $m/\Delta \to 0$ singularities appearing at ${\cal O}(\alpha^2)$.

\subsection{Small-velocity expansion \label{sec:smallbeta}}
Our treatment of $\beta$ as an order unity parameter is justified over the bulk of the neutron decay phase space (cf. Fig.~\ref{fig:betaspec}).
Starting at ${\cal O}(\alpha^3)$, the small-$\beta$ region is no longer kinematically suppressed [for $m/\Delta = O(1)$].  
Here we discuss the third order correction from this region.  

Let us consider the beta decay spectrum, $\dd\Gamma/\dd E$ as the electron energy goes to the electron mass, $E\to m$, or equivalently, $\dd \Gamma/\dd \beta$ as the electron velocity goes to zero, $\beta\to 0$. 
It is well known that photon exchanges with a non-relativistic final-state proton lead to enhancements by powers of $\pi^2/\beta$ leading to contributions of order $(\frac{\alpha}{\pi})^n \times (\frac{\pi^2}{\beta})^n \sim (\frac{\pi \alpha}{\beta})^n$. These terms are captured by the non-relativistic Fermi function and lead to the factorization theorem, 
\begin{equation}\label{NR-factorization}
    \frac{\dd \Gamma}{\dd \beta} \approx  \left(\frac{\dd \Gamma}{\dd \beta} \right)_{\rm tree}\times F_{\rm NR}(\pi \alpha/\beta)~,
\end{equation}
where 
\begin{equation}
    F_{\rm NR}(\pi \alpha/\beta) = \frac{2\pi \alpha/\beta}{1-\exp(-2\pi \alpha/\beta)}~.
\end{equation}
A naive expansion in $\alpha$ of $F_{\rm NR}$ leads to terms whose average over phase space diverges. Yet, when treated non-perturbatively, $\langle F_{\rm NR} \rangle$ is finite. Subtle issues begin at $O(\alpha^3)$ which we will now discuss. 

We begin with the infinite series representation \cite{Melnikov:2014lwa}, 
\begin{equation}\label{MVV}
    F_{\rm NR}(\pi \alpha/\beta) = 1 + \frac{\pi \alpha}{\beta} + \frac{\pi^2 \alpha^2}{3\beta^2}- \frac{2\alpha^2}{\beta^2}\sum_{n=1}^\infty \frac{1}{n^2} \frac{(\alpha/n)^2}{\beta^2 + (\alpha/n)^2}~.
\end{equation}
Integrating over phase space, each of the first three terms yields a finite contribution at $\beta \to 0$.  However, 
 in the remaining series all terms must be retained when $\beta \sim \alpha$, and sum to a contribution that is ${\cal O}(\alpha^3)$. 

Let us consider the contribution to the phase space integral from the region $\beta < \beta_c$ where $\beta_c$ is a cut-off velocity chosen such that $\alpha/\beta_c \sim O(\sqrt{\alpha})$, i.e., $\beta_c$ is small enough that a non-relativistic expansion is justified but large enough that $\alpha/\beta_c$ is still perturbative. In this small-$\beta$ region, we can expand the phase space measure
\begin{equation}
  \dd \Phi \simeq (\Delta-m)^2m^3\beta^2 \dd \beta  + O(\beta^4)~. 
\end{equation}
Let us compute the contribution to the phase space integral from the infinite series in \cref{MVV}:
\begin{equation}
    \label{FF-non-analytic}
  \begin{split}
    (\Delta-m)^2m^3\sum_{n=1}^\infty \frac{1}{n^2} \int_0^{\beta_c} \dd \beta (-2\alpha^2 )
    \frac{(\alpha/n)^2}{\beta^2 + (\alpha/n)^2} &= (\Delta-m)^2m^3|\alpha|^3 \sum_{n=1}^\infty \frac{(-2)}{n^3} \arctan\qty(\frac{n\beta_c}{|\alpha|})\\
    &=(\Delta-m)^2m^3|\alpha|^3 \times (-\pi \zeta_3)  + \ldots~,
    \end{split}
\end{equation}
where in the final line we expanded $\arctan(n\beta_c/|\alpha|)$ about $\beta_c/|\alpha| \to\infty$. Notice that the left-hand side is manifestly even in $\alpha$, requiring the right-hand side to be the even and non-analytic function $|\alpha|^3$. 

At $O(\alpha^3)$, one must also include bound-state contributions from $s$-wave hydrogenic orbitals formed by the $e^-p^+$ final state. 
The contribution is given by the sum over squared wavefunctions evaluated at the origin, with a relative factor of $(2\pi)^3/(4\pi)$ related to 2-body vs. 3-body phase space. The relevant quantity is
\begin{equation}
    \label{BS-non-analytic}
  2\pi^2(\Delta-m)^2\sum_n |\psi_n(\vb{x}=0)|^2 =  2\pi^2(\Delta-m)^2 \sum_n \frac12\left(\alpha^3 + |\alpha|^3\right) \frac{m^3}{\pi n^3} = (\Delta-m)^2m^3 (\alpha^3 + |\alpha|^3) \times(\pi\zeta_3) \,. 
\end{equation}
In particular, when $\alpha <0$ (corresponding to equally charged particles in the final state), the bound state contribution vanishes. 
For both positive and negative values of $\alpha$, the sum of \cref{FF-non-analytic,BS-non-analytic} gives 
\begin{equation}\label{eq:alpha3sum}
    (\Delta-m)^2m^3|\alpha|^3 \times (-\pi \zeta_3)+ (\Delta-m)^2m^3 (\alpha^3 + |\alpha|^3) \times(\pi\zeta_3) = (\Delta-m)^2m^3 \alpha^3  \times(\pi\zeta_3) ~,
\end{equation}
leaving only the analytic-in-$\alpha$ term. This cancellation, and the origin of $|\alpha|^3$ vs. $\alpha^3$,  is discussed in detail in Ref.~\cite{Melnikov:2014lwa} where a similar problem appears in the theory of the electron's anomalous magnetic moment. 

If $\beta_{\rm max}$ (as set by kinematics) were to be very small, $\beta_{\rm max} \sim \alpha$, then the ``large velocity'' region does not exist, one replaces $\beta_c \rightarrow \beta_{\rm max}$ in the above analysis, and the full functional dependence of $\arctan(\alpha/n\beta_{\rm max})$ must be kept in each term of the sum. Such a scenario requires a highly tuned value of $(\Delta-m)$ satisfying $(\Delta-m) \lesssim m \alpha^2$. In this regime, one could imagine that $-\frac12 \alpha^2 m\lesssim (\Delta-m) <0$, in which case continuum contributions would be kinematically forbidden but hydrogenic bound states would be present. 
In practice, such small energy splittings do not occur in Nature and are certainly  not relevant for neutron beta decay.

An alternative formulation describes the impact of Coulomb photon exchange on the inclusive rate via the optical theorem.  For non-relativistic energies, the result is 
\begin{align}\label{eq:disp_int}
    {\rm d}\Phi &\to {\rm d}\Phi \times \bigg[\left( -2\pi \over m^2\beta  \right) {{\rm Im}\, G(E_{\rm NR}) } \bigg] 
    \nl
    &\approx m^3 (\Delta -m)^2 {\rm d}E_{\rm NR}\, 
    \bigg[\left(-2\pi\over m^3\right) {{\rm Im}\, G(E_{\rm NR}) } \bigg]
    \,,
\end{align}
where $E_{\rm NR} = E-m$ and $G$ is given by the Coulomb propagator,
\begin{align}
G(E_{\rm NR}) &= \int{\dd^3p \over (2\pi)^3} \int{\dd^3p^\prime \over (2\pi)^3} \, \langle \vec{p}| {1\over E_{\rm NR} - H_{\rm Coul.} + \iu 0} | \vec{p}^{\,\prime}\rangle \,,
\end{align}
with $H_{\rm Coul.} = p^2/(2m) -\alpha/r$.
A dispersive analysis of $G$ yields (dropping the ``NR" subscript for brevity)
\begin{align}
    G(E) &= {1\over \pi}\int_{-E_0}^\infty \dd E^\prime 
    {{\rm Im}\,G(E^\prime) \over E^\prime - E} \,,
\end{align}
so that 
\begin{align} \label{eq:lim}
    \int_{-E_0}^\infty \dd E^\prime\, {\rm Im}\,G(E^\prime) 
    = \lim_{E\to -\infty}\big[-\pi E G(E) \big] \,, 
\end{align}
where $E_0 = -m\alpha^2/2$ is the ground state energy. 
Subtraction of the $\alpha^0$, $\alpha^1$ and $\alpha^2$ terms in $G$ should be understood, in order that $G$ satisfy an unsubtracted dispersion relation, and so that the limit (\ref{eq:lim}) can be taken.   The perturbative expansion of $G$ at large negative energies is known to all orders in perturbation theory \cite{Melnikov:2014lwa},
\begin{align}
    G - [G^{(0)} + G^{(1)} + G^{(2)}] 
    &= {m^4 \alpha^3 \over \pi p^2 } 
    \bigg[ \zeta_3 + {m\alpha \over \sqrt{-p^2 -\iu 0}}\zeta_4 
    + \left( {m\alpha \over \sqrt{-p^2 -\iu 0}} \right)^2 \zeta_5 + \dots \bigg] \,,
\end{align}
with $p^2 = 2m E_{\rm NR}$.   Integrating Eq.~(\ref{eq:disp_int}) 
over $E_{\rm NR} = -E_0 \dots \infty$, and taking the limit (\ref{eq:lim}),
reproduces the sum of continuum and bound state contributions (\ref{eq:alpha3sum}).

In conclusion, we find that the total decay rate has a well defined expansion in $\alpha$. There are non-perturbative contributions from bound-states and the non-relativistic Fermi function at small electron velocity, however all non-analytic terms cancel for $(\Delta-m)\gg \alpha^2 m$. We have therefore verified that no large enhancements occur from the small-$\beta$ portion of the spectrum, and our analysis presented above is valid through $O(\alpha^2)$ for the physical kinematics of neutron beta decay. .

\subsection{Cancellation of singularities in the small mass expansion \label{sec:KLN}}

We have used the large-energy or equivalently small-mass limit, $m^2/\Delta^2 \ll 1$, 
to evaluate the ``non-enhanced" ${\cal O}(\alpha^2)$ corrections to the hard function
$H(\hat{\mu}^2)$ in Eq.~(\ref{eq:Hratio}).  
Here we comment on apparent violations of the Kinoshita-Lee-Nauenberg~\cite{Lee:1964is,Kinoshita:1962ur} (KLN) theorem which arise due to the kinematic constraint $\Delta < 3 m$. 

In the limit where $m\rightarrow 0$, the KLN theorem  guarantees that the total decay rate will be free of singularities as $m\rightarrow 0$ when expressed in terms of $\overline{\alpha}_1$. 
In the limit $\Delta \gg m$, the total rate is 
\begin{align}
    \Gamma \propto \int_0^\Delta \dd E\, E^2 (\Delta-E)^2 \, 
    \left( \dd \Gamma\over \dd E\right)_{\rm tree} 
    S(\vb*{\epsilon}_\gamma) H(\vb*{\epsilon}_\gamma) \,.
\end{align}
Let us expand the contributions of hard ($H$) and soft ($S$) photons as~\cite{Hill:2023acw} 
\begin{align}
  H(\vb*{\epsilon}_\gamma) 
    &= 1 + {\overline{\alpha}_1 \over 2\pi}\bar{H}^{(1)} 
    +  \left(\overline{\alpha}_1 \over 2\pi\right)^2 \bar{H}^{(2)} + \dots \,,\nl
    S(\vb*{\epsilon}_\gamma) 
    &= 1 + {\overline{\alpha}_1 \over 2\pi}\bar{S}^{(1)} 
    +  \left(\overline{\alpha}_1 \over 2\pi\right)^2 \bar{S}^{(2)} + \dots \,,
\end{align}
where $\bar{H}^{(n)}$ and $\bar{S}^{(n)}$ denote the coefficients in the $\bar{\alpha}_1$ expansion. 
The familiar results at one-loop order are~\cite{Sirlin:1967zza,Hill:2023acw,VanderGriend:2025mdc} 
\begin{align}
    [S^{(1)}_V]_{0} &= 
    \log{\mu \over \lambda} \left( -4L + 4 \right)
    \,, \nl
    [S^{(1)}_R]_{0} &=
    \log{2\vb*{\varepsilon}_\gamma\over \lambda} \left( 4L - 4 \right) - 2L^2 + 2L + 2 - {\pi^2\over 3}
    \,, \nl
    [H^{(1)}_V]_{0} &= 3 \log{\mu_{\rm UV}\over m} + \log{\mu\over m} 
    \left( 4L - 4 \right) 
    - 2L^2 + 2L + {5\pi^2 \over 3} 
    \,, \nl
    [H^{(1)}_R]_{0} &= 
    \log{\vb*{\varepsilon}_\gamma \over \Delta - E}\left( -4L + 4 \right)
    + 2L \bigg[ {(\Delta - E)^2 \over 12 E^2} 
    + {2(\Delta-E)\over 3E} - 3 \bigg]
    - \frac43 {(\Delta-E)\over E} + 6 \,,
\end{align}
where the subscript ``$0$'' indicates the leading term in an expansion in $m/\Delta$ (cf. Ref.~\cite{VanderGriend:2025mdc}).
Integrating over the electron energy, the separate contributions are singular at $m\to 0$,
\begin{align}
    [S^{(1)}]_{0} &\approx -2 \log^2{m} 
    + \log{m} \Big{\langle} -4 \log{\vb*{\epsilon}_\gamma \over E } + 4\log{\mu} - 2 \Big\rangle 
    \,, \nl
    [H^{(1)}_V]_{0} &\approx
    2\log^2{m}
    + \log{m} \left( -4\log{\mu} - 1 \right) 
    \,, \nl
    [H^{(1)}_R]_{0} &\approx 
    \log{m} \Big\langle  4\log{\vb*{\epsilon}_\gamma \over E} +3 \Big\rangle
    \,,
\end{align}
where we write $S^{(1)} = S_V^{(1)} + S_R^{(1)}$, and denote average over phase space by $\langle \cdots \rangle$. 
The electron energy spectrum, $\dd \Gamma/\dd E$, is singular at $m\to 0$, but as expected, the singular contributions cancel when they are added and when the spectrum is integrated to obtain the total rate. 

We remark that the KLN theorem fixes the $\log{m}$ singularity of the one-loop hard real photon contribution
in terms of the soft and hard virtual photon contribution, 
\begin{align}
 [H^{(1)}_R]_{0} \approx  
   - \left(  [S^{(1)}]_{0} +  [H^{(1)}_V]_{0}  \right)
    &\approx 
    \log{m} \Big\langle 4\log{\vb*{\epsilon}_\gamma \over E} + 3 \Big\rangle \,. 
\end{align}
The singularity of the hard real photon contribution 
arises from energetic photons collinear with the electron.  Including photons within an angular cone $\theta_{\gamma} \gg m/\Delta$,~\cite{Tomalak:2021hec,Tomalak:2022xup}
\begin{align}
    [H^{(1)}_{R,\rm collinear}]_{0}
    = 2\int_0^{1-{\vb*{\epsilon}_\gamma \over E}}dx\,
    \bigg[ {1+x^2\over 1-x}\log{x E \theta_{\gamma} \over m}  - {x\over 1-x}\bigg] 
    \approx \Big\langle \log{m\over E \theta_{\gamma}}
    \left( 4 \log{\vb*{\epsilon}_\gamma \over E} +3 \right) \Big\rangle \,. 
\end{align}
The angle $\theta_\gamma$ replaces the electron mass as a regulator for collinear singularities, allowing a smooth $m\to 0$ limit. 

At two-loop order, KLN cancellation of $\log{m}$ singularities in the $m\to 0$ limit requires inclusion of both (single and double) real photon emission, and real additional $e^+e^-$ emission. 
However, physical kinematics dictate that $n\rightarrow p e^+ e^- e^- \overline{\nu}_e$ is forbidden ($\Delta\approx 1.29~{\rm MeV}$ is less than $3 m \approx 1.53~{\rm MeV}$). It follows that there will be effects in the purely virtual corrections (from a vacuum polarization loop) without any compensating real pair-production process.  
To isolate this effect, let us consider the $\log{m}$ singularities which scale with $n_e$: 
\begin{multline}\label{eq:epemlog}
    \Big\langle [\bar{S}^{(2)}]_{0} + [\bar{S}^{(1)}]_{0} [\bar{H}^{(1)}]_{0} + [\bar{H}^{(2)}]_{0}
    \Big\rangle 
    \\ 
    \approx n_e \bigg[ \frac89 \log^3\left({m\over 2\Delta}\right)
    + \frac{344}{45} \log^2\left({m\over 2\Delta}\right)
    + \left( {2371\over 75} - {8\pi^2\over 9} \right)\log\left({m\over 2\Delta}\right) 
   \bigg]  
   + {\cal O}( n_e^0 )
    \,,
\end{multline}
where we have set $\mu=2\Delta$ and performed the phase space average. 
This expression includes the effects of real photon radiation.
The $\log{m}$ singularities in Eq.~(\ref{eq:epemlog}) represent apparent KLN violations due to missing $e^+ e^-$ pair production, as in the physical neutron beta decay process. 
While these contributions are included in our calculation, they are not numerically large, owing to $\log(\Delta/m) \approx 0.9$.  The terms in Eq.~(\ref{eq:epemlog}) contribute to the rate at the level 
$10^{-3} \times ( -0.005 + 0.027 - 0.050 )$. 

\vfill
\pagebreak

\section{Discussion}
Control of QED corrections is an important theoretical challenge for neutron beta decay. Part of this challenge relates to non-perturbative hadronic structure but, at present levels of accuracy, is confined to one-loop order. Long-distance radiative corrections, by way of contrast, are insensitive to hadronic structure but are required at two-loop order. It is well known that these corrections contain certain large numerical enhancements (proportional to $\pi^2$), and their origin has historically been ascribed to the Fermi function (i.e., a resummation of diagrams involving a background Coulomb field). A shortcoming of this argument is that very little of the neutron decay electron spectrum shown in \cref{fig:betaspec} satisfies the non-relativistic kinematics that make the Fermi function a controllable resummation of enhanced contributions. 

In this work we have argued, and explicitly demonstrated, that these numerical enhancements stem from the infrared region of the amplitudes, or the ``soft-hard boundary''.
We have shown that these numerical enhancements can be controlled and resummed using renormalization group methods, making no reference to the non-relativistic limit. 

We have used our results to supply the most precise determination of long-distance radiative corrections to neutron beta decay \cite{VanderGriend:2025mdc}. 
We ascribe an uncertainty for an extraction of $|V_{ud}|$ at the level of $3\times 10^{-5}$ from the long-distance region. 
Our results set a target for uncertainty reductions in the experimental and short-distance inputs, and supply a firm foundation for estimating higher order radiative corrections in processes involving numerically enhanced Coulomb exchanges for both  relativistic and quasi-relativistic kinematics.

\section*{Acknowledgements}
PVG acknowledges support from the Visiting Scholars Award Program of the Universities Research Association.
RP thanks the Institute for Nuclear Theory at the University of Washington for its kind hospitality and stimulating research environment during program INT 23-1b.
This research was supported in part by the INT's U.S. Department of Energy grant No. DE-FG02-00ER41132. Part of the research of RP was performed at the Kavli Institute for Theoretical Physics which is supported by the National Science Foundation under Grant No. NSF PHY-1748958.
RP is supported by the Neutrino Theory Network under Award Number DEAC02-07CH11359, the U.S. Department of Energy, Office of Science, Office of High Energy Physics under Award Number DE-SC0011632, and by the Walter Burke Institute for Theoretical Physics.
This work was supported by the U.S. Department of Energy, Office of Science, Office of High Energy Physics, under Award DE-SC0019095.
This work was produced by Fermi Forward Discovery Group, LLC under Contract No. 89243024CSC000002 with the U.S. Department of Energy, Office of Science, Office of High Energy Physics.

\appendix

\section{Power corrections \label{app:power}}

In Ref.~\cite{VanderGriend:2025mdc}, we have applied the results derived above to compute radiative corrections to the neutron's lifetime. 
In terms of the neutron-proton mass splitting $\Delta=m_n-m_p \simeq 1.3~{\rm MeV}$ 
and the electron mass $m_e = 0.511\,{\rm MeV}$, a natural expansion parameter is 
\begin{align}
\delta^2 &= {m^2\over \Delta^2} = 0.156 \,. 
\end{align}
Here we examine the convergence of this expansion, showing  
that certain terms in the phase space and matrix element must be treated with care. 
The discussion is separated into the phase space itself, the one-loop corrections to the matrix element (where exact results can be used to test the $m/E$ expansion), and the two-loop corrections that are the focus of this work. 

\subsection{Phase space integral}

The rate for neutron beta decay in the static limit is governed by ($\bm{p}$ and $\bm{k}$ are 
the momenta of electron and neutrino, respectively) 
\begin{align}
&\int {d^3 p \over (2\pi)^3 2 E_{\bm p}}  
\int {d^3 k \over (2\pi)^3 2 E_{\bm k}}  
(2\pi) 
\delta( E_{\bm{p}} + E_{\bm{k}} - \Delta) \bar{\Sigma}|{\cal M}|^2
\nl
& \propto {(4\pi)^2\over (2\pi)^5} 
\int_{m}^\Delta dE \, E \sqrt{E^2-m^2} (\Delta - E)^2 
\nl
& = {(4\pi)^2\over (2\pi)^5} \Delta^5 \frac{1}{30}
\bigg[ 1 - 5 \delta^2 + \left( -\frac{15}{8} - \frac{15}{2}\log{\delta\over 2} \right)\delta^4 
+ \frac58 \delta^6 + \frac{5}{128} \delta^8 + \frac{1}{128}\delta^{10} + \dots 
\bigg]
\nl
&= {(4\pi)^2\over (2\pi)^5} \Delta^5 \frac{1}{30}
\bigg[ 1 - 0.78 + 0.25 + 0.0024 + 0.000023 + \dots
\bigg] \,, 
\end{align}
where we use that $\bar{\Sigma}|{\cal M}|^2 \propto 4p^0k^0$. 
The series has large numerical coefficients in the first and second order terms but converges geometrically thereafter (with coefficients $\le 5/8$).
We remark that if one had instead written the integral as a function of $\beta_{\rm max}$ (the maximum electron velocity in the static limit) and expanded as a series in $(\beta_{\rm max})^2$, then it would take $64$ terms in the expansion to achieve permille levels of accuracy. This indicates that a non-relativistic expansion is inappropriate for the problem of neutron beta decay. 

Since the neutron decay phase space has slow convergence properties, we elect~\cite{VanderGriend:2025mdc} to retain the full phase space measure (defined in the static limit) in our averages, whereas the $O(\alpha^2)$ correction to the matrix element is expanded in $m/E$. 
We now turn to results at $O(\alpha)$, which can be used as a testing ground for the convergence of the $m/E$ expansion.

\subsection{One loop correction}

Although the convergence of the $m/E$ expansion is slow for the phase space measure, in this section we will study the convergence of the expansion for the amplitude. Consider the one-loop correction~\cite{Sirlin:1967zza} that includes real radiation, 
\begin{align}\label{eq:sirlinexample}
    \left(\dd\Gamma\over \dd E\right) = 
    \left(\dd\Gamma\over \dd E\right)_{\rm tree}
    \bigg[ 1 + {\alpha\over 2\pi}\left( 3\log{\mu_{\rm UV} \over m} + {2\pi^2\over \beta} + \dots  \right) \bigg] \,.
\end{align}
All $m/E$ dependence lies in the ellipsis in Eq.~(\ref{eq:sirlinexample}) and in $\beta=|\vb{p}|/E = \sqrt{1-m^2/E^2}$. 

Let us first consider the $1/\beta$ term. In the formal $m \rightarrow 0$ limit, we would have $\langle \beta^{-1}-1 \rangle_{m\rightarrow 0} = 0$,
where angle brackets denote phase space average. 
However, the result for physical kinematics differs substantially: $\langle \beta^{-1}-1 \rangle_{\rm phys} = 0.46$. This can be compared to a ``typical'' power correction $\delta^2=0.156$, and suggests that terms with $1/\beta^n$ should be handled with care, especially when they are enhanced by the $X_*^2$ resummation factor discussed in the main text.

Let us expand the remaining radiative correction in powers of $\delta^2=m^2/\Delta^2$, and integrate over the remaining electron phase space ({\it not} expanding the phase space integral).   
Setting $\mu_{\rm UV} = \Delta$ and omitting the term $2\pi^2/\beta$ (which is absorbed in the resummation factor discussed above), the contributions to the total decay rate are
\begin{align}
    1 + \bigg[ (-6.2\times 10^{-3}) + (-1.1\times 10^{-3}) + (2.3 \times 10^{-5}) + \dots  \bigg] \,,
\end{align}
where the terms in square brackets correspond to $\delta^0$, $\delta^2$, $\delta^4$, $\dots$.  

We observe that after resummation (extracting $2\pi^2/\beta$ and setting $\mu_{\rm UV}=\Delta \ll M$), power corrections of order 
$\delta^2$ are of ``natural size'' (i.e. roughly 20\% of the leading correction) and contribute at the permille level to the first order (in $\alpha$) radiative correction. We now turn to two-loop corrections. 

\subsection{Two-loop correction}
For two-loop radiative corrections, we expect a further suppression by $\alpha \sim 10^{-2}$, i.e. power corrections beyond one loop should naively be expected to be phenomenologically unimportant. This expectation is spoiled, however, by the presence of $1/\beta^2$ terms whose phase space average is substantially different from unity, $\langle 1/\beta^2 \rangle_{\rm phys} \approx 2.55$. These terms require special care. 

The enhanced power correction, proportional to 
$1/\beta^2 -1$, is determined by noting that the hard amplitude can be written~\cite{Borah:2024ghn,Plestid:2024eib} 
(cf. Fig.~6 of Ref.~\cite{Borah:2024ghn})
\begin{equation}
    \mathcal{M}_H = \qty[\mathcal{M}_H]_{v\rightarrow -v } + Z \alpha \mathcal{M}_{\rm ext}^{(1)} 
    + Z^2 \alpha^2 \mathcal{M}_{\rm ext}^{(2)} + Z\alpha^2\mathcal{M}_{\rm JR}~ + \order(\alpha^3) \,,
\end{equation}
where $[{\cal M}_H]_{v\to -v}$ and ${\cal M}_{\rm JR}$ are finite at $\beta\to 0$, and 
$\mathcal{M}_{\rm ext}^{(1)}$ and $\mathcal{M}_{\rm ext}^{(2)}$ are 
known analytically with full mass dependence \cite{Hill:2023bfh}. 
Here JR denotes Jaus-Rasche \cite{Jaus:1972hua,Jaus:1986te,Sirlin:1986cc} and $Z$ is a formal counting parameter that must be set equal to $Z=1$ for neutron beta decay. 

The result of this analysis is to add to the spectrum a term $(-1)\times\alpha^2\zeta_2 m^2/(E^2-m^2)$ \cite{VanderGriend:2025mdc}. 
After this correction is included, the small-$\beta$ asymptotics agree with the non-relativistic factorization theorem \cref{NR-factorization}. 
Remaining corrections are finite at $\beta\to 0$, i.e., they are free of $1/\beta$ enhancements; as in the one-loop analysis, we estimate such corrections to be of natural size in the power counting parameter $\delta^2$, and thus small compared to the $10^{-4}$ precision level. 
This expectation is consistent with past studies of the energy dependence of $\mathcal{M}_{\rm JR}$ \cite{Sirlin:1986hpu}.

\section{Hard function from relativistic form factors and UV matching coefficient}\label{app:hard_function}

Here the hard matching coefficient is extracted from relativistic form factors in the large-energy limit. 
Consider the physical form factors for $b$-quark to $u$-quark transitions 
defined by 
\begin{align}
    \langle \bar{u} \gamma^\mu b \rangle 
    = F_1 \bar{u}\gamma^\mu u + F_2 v^\mu \bar{u}u + F_3 {p^\mu \over v\cdot p} \bar{u} u \,. 
\end{align}
For our application, we identify the heavy initial $b$ quark with the anti-proton, and the light final $u$ quark with the electron.  
As usual, we set $T_F=1$, $C_F=1$, $C_A=0$ to obtain QED results from expressions for a general gauge group.   
We track the light flavor dependence by identifying $n_{e}=1$ as the number of dynamical electrons.  
For simplicity, consider the amplitude for Lorentz index $\mu$ such that $v^\mu=0$ and $p^\mu =0$ in the rest frame of the heavy particle.   The amplitude in the full theory is then determined by $F_1$, which we take from Refs.~\cite{Beneke:2008ei,Bonciani:2008wf}.
We work in the limit of infinite heavy-quark mass, i.e., $M\to \infty$.
As noted at the end of Sec.~2 of Ref.~\cite{Beneke:2008ei}, the form factor depends on a harmonic polylogarithm which cannot be expressed as a standard polylogarithm; we evaluate the infinite mass limit here using the HPL \cite{Maitre:2005uu} and PolyLogTools \cite{Duhr:2019tlz} packages for {\it Mathematica}.
The relevant quantity is thus given by
\begin{align}
    F_{1}(M,-E,\mu) =& 
    1+
    \frac{\bar{\alpha}_{1}}{4\pi} \left( -2L_{E}^{2} + 2 L_{E} + \frac{3}{2}L_{M} - 6 - \frac{5\pi^{2}}{12} \right)+
    \left(\frac{\bar{\alpha}_{1}}{4\pi}\right)^{2}\bigg\{n_{e}\bigg[ -\frac{16}{9} L_{E}^{3} + \frac{64}{9} L_{E}^{2} \nonumber \\
    &- \left(\frac{304}{27}+\frac{16\pi^{2}}{9}\right)L_{E} + L_{M}^{2} - 7 L_{M} +\frac{149\pi^{2}}{54} + \frac{2\zeta_{3}}{9} + \frac{6629}{324}\bigg] + 2L_{E}^{4} -4 L_{E}^{3} \nonumber \\
    &+ \left(-3L_{M} +\frac{5\pi^{2}}{6} + 14 \right)L_{E}^{2} + \left( 3L_{M} - \frac{11\pi^{2}}{2} + 24\zeta_{3} -8 \right)L_{E} + \frac{9}{8}L_{M}^{2} \nonumber \\
    &+ \left( -\frac{41}{4} + \frac{17\pi^{2}}{24} \right) L_{M} - \frac{167\pi^{4}}{288} + \frac{2\pi^{2}}{3} + \frac{28\pi^{2}}{3} \log{2}  - 37 \zeta_{3} + \frac{839}{16} \bigg\} + \order(\alpha^3, 1/M) \,.
\end{align}
Here $\bar{\alpha}_1(\mu)$ denotes 
the $\overline{\rm MS}$ coupling in the effective theory with $n_{e}=1$, 
and $L_M \equiv \log(M^2/\mu^2)$. 

We identify the full theory amplitude as a product of a matching coefficient onto heavy particle effective theory, and a matrix element in the low energy theory.  
\begin{align}
    O_{\rm eff} = \bar{u} \Gamma h_v^{(b)} \,.
\end{align}
The ultraviolet matching coefficient is given by Refs.~\cite{Broadhurst:1994se, Grozin:1998kf, Bekavac:2009zc} as
\begin{align}
C(M,\mu) &= 1 + {\bar{\alpha}_1\over 4\pi}\left( 
-4+ \frac32 L_M
\right)
+ \left(\bar{\alpha}_1\over 4\pi\right)^2 
\bigg\{
n_{e} \bigg[ L_M^2  - 7 L_M + {445\over 36} + {2\pi^2\over 3}  \bigg]
\nl
&\quad
+ \frac98 L_M^2 + \left({4\pi^2\over 3} - {29\over 4} \right)L_M 
+ {28\pi^2\over 3}\log{2} 
-{71\pi^2\over 6}
-22\zeta_3 
+ {839\over 16}
\bigg\} \,.
\end{align}
Using dimensional regularization as an infrared regulator, the soft and collinear contributions vanish, leaving 
\begin{align}
     \lim_{M\to \infty}\frac{F_1(M, -E, \mu)}{C(M,\mu)} =  F_H(-E,\mu)\,.
\end{align}

\section{Remainder and jet functions}
The factorization theorem used in the relativistic limit contains components (soft, hard, jet, and remainder functions) that are required through two-loop order.  In this appendix, we discuss the calculation/extraction of the necessary jet and remainder functions.

\subsection{Remainder function \label{app:remainder}}

The remainder function represents a matching between $n_e=1$ and $n_e=0$ 
soft effective theories. Through two loop order, the only nontrivial contribution to the matching is 
the vertex correction, which takes the form 
\begin{align}
    \delta {\cal M} &=  \quad
    \parbox{30mm}{
\begin{fmfgraph*}(100,85)
  \fmfleftn{l}{3}
  \fmfrightn{r}{3}
  \fmftopn{t}{3}
  \fmf{phantom}{l2,v,r2}
  \fmffreeze
  \fmf{double}{r2,x,v}
  \fmf{double}{l2,y,v}
  \fmffreeze
  \fmf{fermion,left,tension=0.25}{a1,a2,a1}
  \fmf{phantom}{t2,a1}
  \fmf{phantom}{t2,a2}
  \fmf{phantom}{x,a1}
  \fmf{phantom}{a2,y}
  \fmffreeze
\fmfv{decor.shape=circle,decor.filled=full,decor.size=2mm}{v}
\fmf{photon,right=0.5}{x,a1}
\fmf{photon,right=0.5}{a2,y}
\end{fmfgraph*} 
}
\nonumber  
    \\[-10mm]  &=
    -\iu e^2 \int {d^d k \over (2\pi)^d}
    {1\over v\cdot k}{1\over v^\prime \cdot k}{1\over (k^2)^2}
    v_\mu v^\prime_\nu \Pi^{\mu\nu}(k) 
    \nl
    &= -\iu e^2 \int {d^d k \over (2\pi)^d}
    \left\langle { v\cdot v^\prime k^2 \over (v\cdot k)(v^\prime \cdot k)}  - 1 \right\rangle  {1\over (k^2)^2} \Pi(k^2)
    \,,
\end{align}
where $\Pi^{\mu\nu}(k) = k^2 g^{\mu\nu} - k^\mu k^\nu$ 
is the one-loop photon polarization tensor obtained from the heavy lepton loop.   
The average in angle brackets is
\begin{align}\label{eq:kav}
     \left\langle { v\cdot v^\prime k^2 \over (v\cdot k)(v^\prime \cdot k)}  - 1 \right\rangle
     &= (2-d) w j(w) -1 \,,
\end{align}
and the remaining integral can be evaluated using the explicit form of $\Pi(k^2)$, 
\begin{multline}
        \Pi(k^2) = {4\iu e^2 \over (d-1) k^2}
    \bigg[ (2-d) \int{d^dL \over (2\pi)^d} {1\over L^2-m^2}
    + \left( {d-2 \over 2}k^2 + 2m^2  \right)
    \int {d^d L \over (2\pi)^d} {1\over L^2 - m^2}{1\over (L+k)^2-m^2}
    \bigg] \,.
\end{multline}
The necessary integral is 
\begin{align}
    -\iu e^2 \int{d^d k \over (2\pi)^d} {1\over (k^2)^2} \Pi(k^2)
    &= {4 e^4 \over d-1}\bigg[ {d-2 \over 2} I_{211} 
    + 2m^2 I_{311} \bigg] \,,
\end{align}
where the integrals $I_{\alpha\beta\gamma}$ are~\cite{Gray:1990yh}
\begin{align}
    I_{\alpha\beta\gamma} 
    &= \int {d^3k \over (2\pi)^d}{d^3l\over (2\pi)^d}{1\over (k^2)^\alpha}
    {1\over [(l+k)^2-m^2]^\beta}
    {1\over (l^2 - m^2)^\gamma}
    \nl
    &= 
    (m^2)^{d-\alpha-\beta-\gamma}
     {(-1)^{\alpha+\beta+\gamma+1}\over 
    (4\pi)^{d}}
    {\Gamma(\alpha+\beta-d/2)\Gamma(\alpha+\gamma-d/2)\Gamma(d/2-\alpha)\Gamma(\alpha+\beta+\gamma-d)
    \over \Gamma(\beta)\Gamma(\gamma)\Gamma(d/2)\Gamma(2\alpha+\beta+\gamma-d)} \,.
\end{align}
Substitution yields 
\begin{align}
 - \iu e^2 (2-d) \int{d^d k \over (2\pi)^d} {1\over (k^2)^2} \Pi(k^2)
 &= \left( {\bar{\alpha}\over 4\pi} \right)^2 \left({m^2\over \mu^2}\right)^{-2\epsilon} 
 \left(
 -{4\over 3\epsilon^2} + {20\over 9\epsilon} - {112\over 27} - {2\pi^2 \over 9}
  + \order(\epsilon)  
 \right)\,.
\end{align}

It is readily seen that the contribution of wavefunction renormalization is given by subtracting the $v^\prime=v$ limit of $\delta{\cal M}$: 
\begin{align}
    \delta Z_h &= {\dd\Sigma \over \dd\omega}\bigg|_{\omega=0}\,,
\end{align}
where 
\begin{align}
    -\iu\Sigma &= - \int{\dd^d k\over (2\pi)^d} {1\over v\cdot k + \omega}{1\over (k^2)^2} v_\mu v_\nu \Pi^{\mu\nu}(k) \,.
\end{align}
The total is thus\footnote{The non-logarithmic term in Eq.~(\ref{eq:kav}) was omitted in Ref.~\cite{Becher:2007cu}, and correspondingly in Ref.~\cite{Hill:2016gdf}, resulting in an unconventional separation between $F_R$ and $F_J$.  The results here correspond to conventional $\overline{\rm MS}$.}
\begin{align}
F_R^{\rm bare} = 1 + 
    \delta M + \delta Z_h 
    &= \big[ (2-d) (w j(w)-1) \big] (-\iu e^2)\int{d^d k \over (2\pi)^d}     {1\over (k^2)^2} \Pi(k^2) \,. 
\end{align}

\subsection{Jet function \label{app:jet-function}}
The product of jet and remainder functions can be obtained from the vector 
form factors for charged particle scattering~\cite{Hill:2016gdf}.  
In the notation of the current paper, the result from Ref.~\cite{Hill:2016gdf} contains the product of two jet functions (since in electron-proton scattering there are two light legs) and is given by, 
\begin{align}
    F_R F_J^2 &= 1 + {\bar{\alpha}_1 \over 4\pi}
    \left(  L_m^2 - L_m + 4 + {\pi^2\over 6} \right)
    + \left({\bar{\alpha}_1 \over 4\pi} \right)^2
    \bigg\{
    n_e\bigg[ L \left( -\frac43 L_m^2-{40\over 9}L_m - {112\over 27} \right) 
    \nl
    &\quad 
    + \frac49 L_m^3 - \frac{14}{9}L_m^2
    + \left( {274\over 27} + {8\pi^2\over 9} \right) L_m
    + {5107\over 162} - {4\zeta_3\over 9} - {41\pi^2\over 27}
    \bigg]
    + \frac12 L_m^4 - L_m^3
     \nl
    &\quad 
    + \left(\frac92 + {\pi^2\over 6} \right)L_m^2
    + \left(-{11\over 2} + {11\pi^2\over 6} -24\zeta_3\right)L_m
    + {241\over 8} + {13\pi^2\over 3} - 8\pi^2\log{2} 
    - 6\zeta_3
    - {163\pi^4\over 360} 
    \bigg\} \,,
\end{align}
which reduces to Eq.~(43) of Ref.~\cite{Hill:2016gdf} at $n_e\to 1$ and $L=\log(2E/m) \to \log(Q^2/m^2)$. 
After dividing by $F_R(\mu)$ from \cref{eq:FRmu} and taking the square root, the two-loop expansion of $F_J(\mu)$ is given by \cref{eq:FJmu}. 
    
\section{Renormalization schemes \label{app:ren-schemes}}

We have worked throughout with dimensional regularization in 
$d=4-2\epsilon$ spacetime dimensions, and have expressed renormalized quantities in the modified minimal subtraction ($\overline{\rm MS}$) renormalization scheme.  In this scheme, the renormalized coupling is defined to subtract commonly-occurring factors of  $\frac{1}{\epsilon} + \log{4\pi} - \gamma_{\rm E}$.
This is achieved by defining the renormalized coupling 
\begin{equation}\label{eq:MSbaralpha}
    \overline{\alpha}(\mu) = Z_{\alpha}^{-1} {e_{\rm bare}^2 \over 4\pi} \left( \frac{4\pi}{\mu^{2}} {\e}^{-\gamma_{\rm E}}\right)^{\epsilon}\,, 
\end{equation}
in terms of the bare Lagrangian parameter $e_{\rm bare}$. 
The coupling renormalization factor $Z_\alpha$, and other operator renormalization factors, are specified in the $\overline{\rm MS}$ scheme to contain only negative powers of $\epsilon$ at each order in an expansion in $\bar{\alpha}$. 

The precise choice of factors in Eq.~(\ref{eq:MSbaralpha}) do not impact physical observables, but do impact the conventional separation between different components of factorized expressions.
As an example, we denote by $\overline{\rm MS}^\prime$ the scheme where 
\begin{equation}\label{eq:MSbaralpha}
    \overline{\alpha}_{\overline{\rm MS}^\prime} (\mu) = Z_{\alpha,\overline{\rm MS}^\prime}^{-1} {e_{\rm bare}^2 \over 4\pi} \left( \frac{4\pi}{\mu^{2}} \Gamma(1+\epsilon) \right)^{\epsilon}\,. 
\end{equation}
In the $\overline{\rm MS}^\prime$ scheme we find instead of Eq.~(\ref{eq:FH}), (using that $\bar{\alpha}=\bar{\alpha}^\prime$ at $\epsilon=0$) 
\begin{align}
    F_H^{(1)}(\hat{\mu}^2)|_{\overline{\rm MS}^\prime} &= - 2 - 2\zeta_2
    \,,
    \nl
    F_H^{(2)}(\hat{\mu}^2)|_{\overline{\rm MS}^\prime}
    &= n_{e}\bigg[ 
    {656\over 81} + 12 \zeta_2
    \bigg] 
    - 8 + 64 \zeta_2 - 15 \zeta_3 - 55\zeta_4
    \,,
\end{align}
leading to [cf. Eq.~(\ref{order-1-numbers})]
\begin{align}
        \big( |F_H(\hat{\mu}^2)|^2\big)|_{\overline{\rm MS}^\prime} &= 1 - 2.6 \left({\bar{\alpha}\over \pi}\right) + (3.5+ 4.2 n_e) \left(\bar{\alpha}\over \pi\right)^2 + \dots \,.
\end{align}

We remark that a subtraction scheme commonly used in chiral perturbation theory \cite{Gasser:1983yg,Ando:2004rk, Cirigliano:2023fnz,Cirigliano:2024msg}, denoted $\overline{\rm MS}_{\chi}$, defines
\begin{equation}\label{eq:MSbaralpha}
    \overline{\alpha}_\chi (\mu) = Z_{\alpha,\chi}^{-1} {e_{\rm bare}^2 \over 4\pi} \left( \frac{4\pi}{\mu^{2}} \e^{-\gamma_E + 1}\right)^{\epsilon}\,. 
\end{equation}
When relating quantities computed in $\overline{\rm MS}$ and $\overline{\rm MS}_\chi$, it is useful to note that (at $\epsilon=0$), 
    $\overline{\alpha}_\chi (\mu) = \overline{\alpha}( \mu/\sqrt{\e} )$, where $\e=\exp(1)\approx 2.718$. 
\end{fmffile}

\bibliography{largepi}

\end{document}